\documentclass[aps, prd, reprint, groupedaddress, amsmath, amssymb]{revtex4-1}
\usepackage{graphicx}
\usepackage{mathrsfs}
\usepackage{bm}
\usepackage{amssymb}
\usepackage{amsmath}
\usepackage{epsfig}
\usepackage{color}
\usepackage{mathtools}
\usepackage{cases}
\usepackage{booktabs}

\usepackage[
colorlinks=true, 
filecolor=black, 
anchorcolor=blue, 
linkcolor=blue, 
citecolor=cyan, 
urlcolor=cyan, 
linktocpage=true, 
plainpages=false, 
breaklinks=true, 
pdfstartview=FitH
]{hyperref}

\def\aCS{ \alpha}   
\def\fa{{[a]}}
\def\fb{{[b]}}
\def\fm{U}

\def\dcs{\omega}
\def\vtheta{\vartheta}
\def\scalar{\varphi}
\def\di{[\mathrm{D}]}
\def\da{[\mathrm{a}]}
\def\db{[\mathrm{b}]}
\def\kk{(\mathrm{K})}
\def\dn{{D}}
\def\dd{\mathrm{d}}
\def\ii{\mathrm{i}}
\def\ee{{e}}

\begin{document}

\title{Slowly rotating charges from Weyl double copy for Kerr black hole with Chern-Simons correction}

\author{Yi-Ran Liu$^{1,3,4}$}
\email
{liuyiran211@mails.ucas.ac.cn}
\author{Jing-Rui Zhang$^{1,3,4}$}
\email
{zhangjingrui22@mails.ucas.ac.cn}
\author{Yun-Long Zhang$^{2,1,5}$}
\email
{zhangyunlong@nao.cas.cn}
\affiliation{$^{1}$ School of Fundamental Physics and Mathematical Sciences,  Hangzhou   Institute for Advanced Study, UCAS, Hangzhou 310024, China. }
\affiliation{$^{2}$ National Astronomical Observatories, Chinese Academy of Sciences, Beijing, 100101, China}
\affiliation{$^{3}$CAS Key Laboratory of Theoretical Physics, Institute of Theoretical Physics, Chinese Academy of Sciences, Beijing 100190, China.} 
\affiliation{$^{4}$University of Chinese Academy of Sciences, Beijing 100149, China.}
\affiliation{$^{5}$ International Center for Theoretical Physics Asia-Pacific, Beijing/Hangzhou, China
}

\date{\today}

\begin{abstract}
The Weyl double copy builds the relation between gauge theory and gravity theory, especially the correspondence between gauge solutions and gravity solutions. In this paper, we obtain the slowly rotating charge solutions from the Weyl double copy for the Kerr black hole with small Chern-Simons correction. Based on the Weyl double copy relation, for the Petrov type D solution in Chern-Simons modified gravity, we find the additional correction to the electromagnetic field strength tensor of the rotating charge. For the Petrov type I solution, we find that the additional electromagnetic field strength tensors have external sources, while the total sources vanish at the leading order.
\end{abstract}

\maketitle


\tableofcontents
\allowdisplaybreaks

\section{Introduction}
\label{section:1}  
The double copy as a rather old topic has been first studied in string theory, then found in a now frontier realm of scattering amplitudes in quantum field theory, such as the Kawai–Lewellen–Tye (KLT) relation~\cite{Kawai:1985xq} and the Bern-Carrasco-Johansen (BCJ) relation~\cite{Bern:2008qj, Bern:2010ue}. Intrigued by the successful structure of double copy in scattering amplitudes, its classical counterpart of field equations has been investigated with some attempts given in ~\cite{Bern:1999ji, Saotome:2012vy, Neill:2013wsa}. Since the double copy has always been perturbative in quantum field theory, it is particularly appealing that exact relations between the classical solutions of gauge and gravity theory make sense, see e.g. as  the double copy and black hole solutions in~\cite{Monteiro:2014cda}. A succinct review for recent progress can be found in~\cite{Adamo:2022dcm}.  

Preliminary attempts on classical version of double copy have been put on the certain class of solutions that linearise the classical field equations.
Kerr-Schild metrics belong to this class, and so do certain multi-Kerr-Schild metrics. The Kerr-Schild double copy has been investigated in~\cite{Monteiro:2014cda}, with an interesting applications in the Kerr-Taub-NUT solution~\cite{Luna:2015paa}. Also, the Eguchi-Hanson metric was studied in
this context using the Kerr-Schild double copy~\cite{Berman:2018hwd}. More recent discussions about Kerr-Schild double copy can be found in~\cite{Alkac:2022tvc, Dempsey:2022sls, Easson:2023dbk, Farnsworth:2023mff, Ortaggio:2023cdz, Ceresole:2023wxg}. 

In order to derive a more generic  exact double copy, and inspired by the Petrov classification \cite{Stephani:2003tm}, the Weyl double copy was invented that treats solutions of gauge fields as the ``single copy". The original context of Weyl double copy deals with Petrov type D solutions of Einstein equation and discuss the plausibility of double copy being extended to more general type of spacetime~\cite{Luna:2018dpt}. Some other equivalent interpretations of double copy have been established with various methods~\cite{Luna:2016due, Goldberger:2016iau, Luna:2016hge, Arkani-Hamed:2019ymq, Huang:2019cja, Kim:2019jwm, Luna:2020adi, Cristofoli:2020hnk, Monteiro:2020plf, Guevara:2020xjx}, as well as in the perturbative theories~\cite{Cardoso:2016ngt, Cardoso:2016amd, Cheung:2016say, Cheung:2017kzx, Goldberger:2017frp, Luna:2017dtq, Goldberger:2017vcg, Chester:2017vcz, Goldberger:2017ogt, Li:2018qap, Shen:2018ebu, Plefka:2018dpa, Mizera:2018jbh, CarrilloGonzalez:2018ejf}.
The Weyl double copy has been investigated in the fluid/gravity duality \cite{Cai:2013uye,Keeler:2020rcv} and in the gravitational waves \cite{Godazgar:2020zbv}.
A twistorial foundation of Weyl double copy has also been proposed in linearized level~\cite{White:2020sfn, Chacon:2021wbr}, which has been applied in topological massive gravity and massive gauge theory in three dimension~\cite{CarrilloGonzalez:2022ggn}. Some substances on double copy on curved space have been discussed in~\cite{Han:2022ubu, Han:2022mze, Bahjat-Abbas:2017htu}, and the hidden symmetry of double copy was given in~\cite{Ball:2023xnr}. The asymptotic formula of Weyl double copy can be found in~\cite{Godazgar:2021iae}, as well as the Newman-Penrose version of that \cite{Mao:2023yle}. The classical double copy has become an active area, and different aspects of classical double copy has aroused many interests~\cite{Easson:2020esh, Chawla:2023bsu, Lin:2022jrp, Lin:2023rwe, Alkac:2023glx, He:2023iew}.

In this paper, we study the Weyl double copy for slowly rotating black hole in Chern-Simons modified gravity. For the Kerr black hole solution, there is a concise and fully detailed context in the reference~\cite{Adamo:2014baa}. 
The Chern-Simons modified gravity has aroused many interests, and applied to cosmology and astrophysics~ \cite{Alexander:2009tp,Yunes:2009hc}.
It is a modified Einstein gravity, and the equations derived from its Lagrangian is not in vacuum, but carrying a source. Thus, to study its single copy, the need for a sourced double copy is self-evident. A primary discussion on sourced Weyl double copy was given in~\cite{Easson:2021asd}, and see some later progress in~\cite{Easson:2022zoh}. 
We will use the perturbative solutions of slowly rotating black holes in Chern-Simons modified gravity~\cite{Yagi:2012ya}, and the Petrov type solution provided in~\cite{Owen:2021eez}.

This paper is organized as follows:
section \ref{section:2} is a overview of Weyl double copy for Kerr black hole and a brief introduction to Chern-Simons modified gravity. 
We derive the Weyl double copy for Petrov type D solution of Slowly rotating black hole in Chern-Simons modified gravity in section \ref{section:3}, and for Petrov type I solution of that in section \ref{section:4}.
We make the conclusion and discussion in section \ref{section:5}.

\section{Notation and setup}
\label{section:2}

All throughout, we take the geometry units: $c=1$ and $G=1$ and stick to the following conventions:   the Greek letters  $\{\mu,\nu,...\}$ stand for the indices of curved spacetime, lowercase Latin letters $\{a,b,...\}$ are denoted as flat indices, and the metric signature is $(-1,1,1,1)$. Capital Latin letters $\{A,B,...\}$ represent spinor indices.
Some of the notations are base on the solution classification in~\cite{Stephani:2003tm}, and spinor techniques in~\cite{Penrose:1985bww, Penrose:1986ca, Penrose:1960eq}.

\subsection{Newman-Penrose formulation}
In the Newman-Penrose formalism, the metric can be decomposed into four null tetrad vectors
\begin{equation}\label{metrick}
    g_{\mu\nu}=-2l_{(\mu}n_{\nu)}+2m_{(\mu}\overline{m}_{\nu)}.
\end{equation}
The null tetrad vectors are normalised as
\begin{equation}
\begin{aligned}
l_{\mu} n^{\mu}=-1,\quad
m_{\mu} \overline{m}^{\mu}=1,
\end{aligned}
\end{equation}
while the rest of the products vanish. The corresponding orthogonal vierbein formalism $\ee^a_{\mu}$ reads
\begin{equation}
  \begin{aligned}
    \ee^{0}_{\mu}&=\frac{1}{\sqrt{2}}(n_{\mu}+l_{\mu}),\ \ \quad
     \ee^{1}_{\mu}=\frac{1}{\sqrt{2}}(m_{\mu}+\overline{m}_{\mu}),\\
      \ee^{2}_{\mu}&=-\frac{\ii}{\sqrt{2}}(m_{\mu}-\overline{m}_{\mu}),\quad
       \ee^{3}_{\mu}=\frac{1}{\sqrt{2}}(l_{\mu}-n_{\mu}).
\end{aligned}
\end{equation}
The metric in \eqref{metrick} can be rewritten as
\begin{equation}
\begin{aligned}
    g_{\mu\nu}&=\ee^a_{\mu}\ee^b_{\nu}\eta_{ab},\quad
    \eta_{ab}=\text{diag}(-1,1,1,1).
\end{aligned}
\end{equation}
The associated frame tetrad
 $l^{a}=\ee^{a}_{\mu}l^{\mu}$.
\begin{equation}\label{metric0}
    \eta_{ab}=-2l_{(a}n_{b)}+2m_{(a}\overline{m}_{b)}.
\end{equation}
We will work with the tetrad set
\begin{equation}
  \begin{aligned}
l_a&=\frac{1}{\sqrt{2}}(1,0,0,1),\ \ \qquad
m_a=\frac{1}{\sqrt{2}}(0,1,\ii,0),\\
n_a & =\frac{1}{\sqrt{2}}(1,0,0,-1),\qquad
\overline{m}_a=\frac{1}{\sqrt{2}}(0,1,-\ii,0).
\end{aligned}
\end{equation}

\subsection{Spinor formulation}\label{spinor formula}
For a spinor basis $\{o_{A},\iota_{A}\}$, the indices are raised and lowered by the two-dimensional full anti-symmetric tensor
\begin{equation}
 \varepsilon^{AB}=
\begin{pmatrix}
      0 & 1\\
     -1 & 0
\end{pmatrix}
=-\varepsilon_{AB}.
\end{equation}
The Infeld-Van der Waerden symbol is
\begin{equation}
\begin{aligned}
\sigma_{AB}^{ab}&=\sigma_{A\dot{A}}^{[a}\bar{\sigma}^{b]\dot{A}C}\varepsilon_{CB},\\
\end{aligned}
\end{equation}
where 
\begin{gather}
\bar{\sigma}^{\mu\dot{A}A}=(\ee^{-1})_{a}^{\mu}\bar{\sigma}^{a\dot{A}A},\\
\sigma^a\equiv\frac1{\sqrt{2}}(1,\vec{\sigma}),\quad\bar{\sigma}^{a }=\frac{1}{\sqrt{2}}(1,-\vec{\sigma}),
\end{gather}
and $\vec{\sigma}$ are the Pauli matrices.
The Weyl spinor and Maxwell spinor can be written as
\begin{align}
    \Psi_{ABCD}&=\frac{1}{4}C_{\mu\nu\rho\lambda}\sigma_{AB}^{\mu\nu}\sigma_{CD}^{\rho\lambda},\label{spinorT1}\\
    f_{AB}&=\frac{1}{2}F_{\mu\nu}\sigma_{AB}^{\mu\nu}.\label{spinorT2}
\end{align}  
What we will utilize here is 
\begin{equation}\label{spinor CF}
\begin{aligned}
   C_{A\dot{A}B\dot{B}C\dot{C}D\dot{D}}&=\Psi_{ABCD}\varepsilon_{\dot{A}\dot{B}}\varepsilon_{\dot{C}\dot{D}}+\bar{\Psi}_{\dot{A}\dot{B}\dot{C}\dot{D}}\varepsilon_{AB}\varepsilon_{CD},\\
   F_{A\dot AB\dot B}&=f_{AB}\varepsilon_{\dot A\dot B}+\bar f_{\dot A\dot B}\varepsilon_{AB}.
\end{aligned}
\end{equation}
where $\bar{\Psi}$ are the complex conjugate of $\Psi$,
and the left-hand side is the result of tensors with each indices contracted with $\sigma_{A\dot{A}}^\mu$. 
One can always use the inverse vierbein to contract with the analogy of tensors \eqref{spinor CF} to get the pure tensor form.

We make use of the notations from \cite{Easson:2022zoh} but slightly different from their choice because of the vierbein deformation here. The basis are associated with frame tetrad by
\begin{align}
\begin{aligned}
    o_A\overline{o}_{\dot{A}}&=l_a\sigma^a_{A\dot{A}},\qquad \iota_A
    \overline{\iota}_{\dot{A}}=n_a\sigma^a_{A\dot{A}},\\
     o_A\overline{\iota}_{\dot{A}}&=m_a\sigma^a_{A\dot{A}},\quad\iota_A\overline{o}_{\dot{A}}=\overline{m}_a\sigma^a_{A\dot{A}}.
\end{aligned}
\end{align}
The spinor basis satisfies the normalization: $o^A\iota_A=1=-\iota^Ao_A$,
we then have the basis
\begin{align}
\begin{aligned}
    o_A&=(1,0),\quad
    \iota_A=(0,1).
    \end{aligned}
\end{align}
A quick look at the definition of Weyl scalars:
\begin{equation}
\begin{aligned}\label{Weylscalars}
\Psi_{0}&=\Psi_{ABCD}o^Ao^Bo^Co^D=C_{\mu\nu\rho\lambda} l^\mu m^\nu l^\rho m^\lambda, \\
\Psi_{1}&=\Psi_{ABCD} o^A o^Bo^C\iota^D=C_{\mu\nu\rho\lambda} l^\mu n^\nu l^\rho m^\lambda, \\
\Psi_{2}&=\Psi_{ABCD}o^Ao^B\iota^C\iota^D=C_{\mu\nu\rho\lambda} l^\mu m^\nu\bar{m}^\rho n^\lambda, \\
\Psi_{3}&=\Psi_{ABCD}o^A\iota^B\iota^C\iota^D=C_{\mu\nu\rho\lambda} l^\mu n^\nu \bar{m}^\rho n^\lambda, \\
\Psi_{4}&=\Psi_{ABCD}\iota^A\iota^B\iota^C\iota^D=C_{\mu\nu\rho\lambda} n^\nu \bar{m}^\nu n^\rho\bar{m}^\lambda .
\end{aligned}
\end{equation}
The most general Weyl spinor can be decomposed into
\begin{align}\label{Weylspinor}
  \Psi_{ABCD}
=&\Psi_{0}\iota_A\iota_B\iota_C\iota_D-4\Psi_1o_{(A}\iota_B\iota_C\iota_{C)}+6\Psi_2o_{(A}o_B\iota_C\iota_{D)}  \nonumber\\
&-4\Psi_3o_{(A}o_Bo_C\iota_{D)}+\Psi_4o_Ao_Bo_Co_D.  
\end{align}

\subsection{Weyl double copy for Kerr BH}\label{weyl for kerr bh }

The key idea of Weyl double copy is a square relation between exact solutions in gravity and in (flat spacetime) gauge theory \cite{Luna:2018dpt}, which was written as
\begin{equation}
{\Psi_{ABCD} = \frac{1}{\scalar_0}\, f_{(AB}\, f_{CD)}}\label{doublecopy}.\
\end{equation}
For the Kerr black hole, one can derive the spinor form of Weyl tensor using the Kinnersley tetrad
\begin{align}
&l^\mu_\mathrm{GR}\partial_\mu~= \frac{r^2+a^2}{\Delta}\partial_t+\partial_r+\frac{a}{\Delta}\partial_\phi , \nonumber\\
&n^\mu_\mathrm{GR}\partial_\mu=  \frac{(r^2+a^2)}{2\Sigma}\partial_t-\frac{\Delta}{2\Sigma}\partial_r+\frac{a}{2\Sigma}\partial_\phi,\nonumber\\
&m^\mu_\mathrm{GR}\partial_\mu= \frac{{\ii}a\sin\theta\partial_t+\partial_\theta+{\ii} \csc\theta\partial_\phi}{\sqrt{2}(r+{\ii} a\cos\theta)} ,
\end{align}
where $\Sigma=r^2+a^2\cos^2{\theta}$ and $\Delta=r^2-2Mr+a^2$. Then, the Weyl spinor for Kerr spacetime is
      \begin{equation}
\begin{split} 
    \Psi_{ABCD}&=-\frac{6M}{(r-{\ii}  a\cos\theta)^3}o_{(A}o_B\iota_C\iota_{D)}.
    \end{split}
\end{equation}

From the Weyl double copy relation \eqref{doublecopy}, the scalar field $\scalar_0$ is set as the zeroth copy 
\begin{equation}
\scalar_{0}=-\frac{q^2}{6M}\frac{1}{r-{\ii}  a\cos\theta},
\end{equation}
which satisfies the vacuum equation of motion
$\dn^\mu\dn_\mu\scalar=0$.
We use $\nabla_\mu$ to denote the covariant derivative in the gravitational spacetime, and $\dn_\mu$ to denote the covariant derivative in flat spacetime with rotation.
Note that the ``$r$'' here is related with the ordinary radial coordinate $r$ through the rotation \cite{Newman:1965tw}.
Then the corresponding Maxwell spinor as the single copy can be obtained as \begin{equation}\label{sc}
f^{}_{AB}=\frac{q \, o_{(A}\iota_{B)}}{(r-{\ii}  a\cos\theta)^2},
\end{equation}
which, after being converted into tensor form, satisfies vacuum Maxwell equations $\dn^\mu F_{\mu\nu}=0$.

\subsection{Chern-Simons modified gravity}
The Chern-Simons modified gravity is a fascinating topic in gravity and cosmology as reviewed in~\cite{Alexander:2009tp, Yunes:2009hc}.
Here we consider the action 
\begin{align}
 S  =~\int \dd^4x\sqrt{-g}\left(\kappa R- \frac{1}{2} \nabla_\mu{\vtheta}\nabla^\mu{\vtheta}  +\frac{\aCS}{4} {\vtheta}R\tilde{R}  \right),
\end{align}
where the Pontryagin density $R\tilde{R} \equiv {R}_{\mu\nu\rho\sigma}\tilde{R}^{\mu\nu\rho\sigma}$,
and
$\tilde{R}^{\mu\nu\rho\sigma}\equiv\frac{1}{2}\varepsilon^{\rho\sigma\alpha\beta}R^{\mu\nu}_{\quad\alpha\beta}$. We consider dynamical framework of the field $\vtheta$, which leads to non-vanishing stress energy tensor
\begin{equation}
T^{({\vtheta})}_{\mu\nu}= \nabla_\mu{\vtheta}\nabla_\nu{\vtheta}-\frac{1}{2}g_{\mu\nu}\nabla_\rho{\vtheta}\nabla^\rho{\vtheta}.
\end{equation}
The equations of motion are
\begin{equation}
\begin{aligned}
&R_{\mu\nu}-\frac{1}{2}g_{\mu\nu}R+\frac{\alpha}{\kappa}C_{\mu\nu}=\frac{1}{2\kappa}{T}^{({{\vtheta}})}_{\mu\nu}, \\
&\nabla^\mu\nabla_\mu{\vtheta}=-\frac{\alpha}{4}R\tilde{R} .
\end{aligned}
\end{equation}
Here, $C^{\mu\nu}$ is called C-tensor, given by:
\begin{equation}
C^{\mu\nu}=(\nabla_\rho{\vtheta})\epsilon^{\rho \sigma \lambda(\mu}\nabla_\lambda R^{\nu)}_{\ \  \sigma}+(\nabla_\rho\nabla_\sigma{\vtheta})\widetilde{R}^{\rho(\mu\nu)\sigma}.
\end{equation}

\section{Petrov type D solution}
\label{section:3}

In this section, we will showcase our work for type D black hole solutions in Chern-Simons modified gravity. 
By inserting the slow-rotation limit, Yunes and Pretorius found the perturbative black hole solution in the full dynamical framework \cite{Yunes:2009hc}, the leading order solutions for the metric and field $\vtheta$ are
\begin{align} \label{typeDmetric}
{\dd}s^{2} &= {\dd}s^{2}_{\kk}  +
2 \aCS\dcs r \sin^{2}{\theta} {\dd}t {\dd}\phi, 
 \\
{\vtheta} &=  \frac{5 \aCS}{8} \frac{a}{M} \frac{\cos{\theta}}{r^2} \left(1 + \frac{2 M}{r} + \frac{18 M^2}{5 r^2} \right),
\end{align}
in which 
\begin{align}
\dcs=\frac{5\aCS}{8} \frac{a}{\kappa r^5}
    \left(
    1+\frac{12}{7}\frac{M}{r}+\frac{27}{10}\frac{M^2}{r^2}
    \right).
\end{align}
${\dd}s^{2}_{\kk}$ means slow-rotation limit of the Kerr metric. Up to order ${\cal O}(\alpha^2 a )$ and ${\cal O}(a^2)$, the non-vanishing components of the metric are
\begin{align}
\begin{aligned}
&g_{tt} = -\fm  - \frac{2M a^{2}}{r^{3}} \cos^{2}{\theta},
\\
&g_{t\phi}  = - \frac{2 M a}{r} \sin^{2}{\theta} + {\aCS}\dcs r\sin^2{\theta},
 \\
&g_{rr}  = \frac{1}{\fm} + \frac{a^{2}}{\fm r^{2}} \left(\cos^{2}{\theta} - \frac{1}{\fm} \right), 
 \\
&g_{\theta \theta}  = r^{2} + a^{2} \cos^{2}{\theta}, 
 \\
&g_{\phi \phi} = r^{2} \sin^{2}{\theta} + a^{2} \sin^{2}{\theta} \left(1 + \frac{2 M}{r} \sin^{2}{\theta} \right), 
\end{aligned}
\end{align}
where
\begin{align}
\fm &\equiv 1-\frac{2M}{r}.
\end{align}

Now we use a tetrad, which is similar to  the Kinnersley tetrad, but with some adjustments to the metric in \eqref{typeDmetric}, and accurate up to $\mathcal{O}(\alpha^2 a )$ and $\mathcal{O}(a^2)$. 
\begin{align}
\begin{aligned}\label{tetradD0}
l^\mu\partial_\mu~=&~l^\mu_\mathrm{GR}\partial_\mu -\aCS \left(\frac{\dcs}{{\fm} r} + \frac{r^2\dcs''}{12M}\right)\partial_\phi , \\
n^\mu \partial_\mu~=&~n^\mu_\mathrm{GR}\partial_\mu  -\aCS \frac{\fm}{2}\left(\frac{\dcs}{{\fm} r} + \frac{r^2\dcs''}{12M}\right)\partial_\phi,\\
m^\mu\partial_\mu=&~m^\mu_\mathrm{GR}\partial_\mu - \ii\aCS 
\frac{ r\sin\theta}{\sqrt{2}}
\frac{r^2\dcs''}{12 M}\partial_t.
\end{aligned}
\end{align}
Here we utilize the null rotation transformation in Appendix \ref{Appa}. The resulting total symmetric Weyl spinor reads
\begin{align}\label{WeylD}
    \ \Psi_{ABCD}^{\di}
=& -\frac{6M}{(r-{\ii}  a\cos\theta)^3}\left(1-{\ii}\frac{3\alpha }{ r^2}\frac{\vtheta}{\kappa}\right)o_{(A}o_B\iota_C\iota_{D)},
\end{align}
which turns out that the spacetime is Petrov type D at this order. The identical relation
\begin{align}\label{thetaomega}
    \vtheta&=-\frac{\kappa r^5\cos\theta}{6M}\left(\frac{\dcs}{r}\right)'
\end{align}
is also used to represent the results in a compact form, where “ $'$ ” means the partial derivative respect to $r$. 
In the following, we will use two kinds of choices for the zeroth copy (scalar fields $\scalar_{\da}$ and $\scalar_{\db}$), and decompose the Weyl spinor \eqref{WeylD} into the gauge fields as:
\begin{equation}\label{typeD}
{\Psi_{ABCD}^{\di} = \frac{1}{\scalar_{\da} }\, f^{\da} _{(AB}\, f^{\da} _{CD)}}
= \frac{1}{\scalar_{\db} }\, f^{\db} _{(AB}\, f^{\db} _{CD)}.
\end{equation}
In the calculation for the Petrov type D case, we work up to order ${\cal O}(\alpha^2 a )$ and ${\cal O}(a^2)$.

\subsection{Decomposition I}

By setting the scalar field in \eqref{typeD} as the zeroth copy
\begin{align}
    \scalar_{\da}=  \scalar_{0}=-\frac{q^2}{6M}\frac{1}{r-{\ii}  a\cos\theta},
\end{align}
the Weyl spinor \eqref{WeylD} can be written as the square of the single copy $f_{AB}$, which can be solved as
\begin{align}
    f_{AB}^{\da}=&
    \frac{q}{(r-{\ii}  a\cos\theta)^2}\left(1-{\ii}\frac{3\alpha }{2 r^2}\frac{\vtheta}{\kappa}\right)o_{(A}\iota_{B)}.
\end{align}
Therefore, we can derive the slowly rotating charges with sourced electromagnetic field from the Weyl double copy formula for the Petrov type-D solution \eqref{typeDmetric}. The corresponding Maxwell field strength tensor is corrected as
\begin{equation}
    F_{\mu\nu}^{\da}= F^{\kk}_{\mu\nu}+ F^{(\aCS)}_{\mu \nu},
\end{equation}
where the Maxwell field strength tensor  $F^{\kk}_{\mu\nu}$ represents the single copy of the Kerr black hole, with non-vanishing components up to $\mathcal{O}(a^2)$,
\begin{equation}
\begin{aligned}
F^{\kk}_{tr} &= -\frac{ q}{r^2}\left(1 -\frac{3a^2 \cos^2\theta}{r^2}\right),    \\
F^{\kk}_{r\phi}  &= -\frac{qa\sin^2\theta}{r^2},\\
F^{\kk}_{t\theta} &= \frac{qa^2\sin2\theta}{r^3},\\  
F^{\kk}_{\theta\phi}& =\frac{qa\sin2\theta}{r}.
\end{aligned}
\end{equation}
The resulting field strength tensor is also consistent with the gauge potential in reference \cite{Monteiro:2014cda}, up to the constant coefficient. The non-vanishing components of the correction $F_{\mu\nu}^{(\alpha)}$ due to Chern-Simon term are
\begin{equation}
\begin{aligned}
    F^{(\aCS)}_{r\phi} &= \frac{{\aCS}q \sin^2\theta }{12M} {r^2\dcs''},\\
    F^{(\aCS)}_{\theta\phi} &=-\frac{3{\aCS} q\sin\theta}{2 r^2}\frac{\vtheta}{\kappa}.
\end{aligned}
\end{equation}

The corresponding equations of motion for the scalar field and Maxwell field are
\begin{align}
&\dn^{\mu}\dn_{\mu}\scalar_{\da}=0,\\
&\dn^\mu F^{\da}_{\mu\nu}\equiv J^{\da}_{\nu},
\end{align}
where the source current is calculated to be
\begin{gather}\label{current0}
J^{\da}_{\nu}=\left(0,0,0,  \frac{{\aCS}q\sin^2\theta}{12 M} \left[\frac{18M}{r^4 \cos\theta}\frac{\vtheta}{\kappa}+ (r^2 {\dcs}'')'\right]  \right).
\end{gather}
With the using of the identical relation \eqref{thetaomega}, the non-vanishing component in \eqref{current0} can also be written as
 \begin{gather}
    J^{\da}_{\phi}= \frac{{\aCS}q\sin^2\theta}{12 M} \left[-3r\left(\frac{\dcs}{r}\right)'+ (r^2 {\dcs}'')'\right]. 
 \end{gather}

\subsection{Decomposition II}

By setting the scalar field in \eqref{typeD} as
\begin{align}
    \scalar_{\db}=-\frac{q^2}{6M}\frac{1}{r-{\ii}  a\cos\theta}\left(1+{\ii} \frac{3\alpha }{ r^2} \frac{\vtheta}{\kappa}\right),
\end{align}
then the Maxwell spinor as the single copy can be \begin{align}
    f_{AB}^{\db}=\frac{q}{(r-{\ii}  a\cos\theta)^2}o_{(A}\iota_{B)}.
\end{align}
Notice that, components of $f_{AB}^{\db}$ equal to that in \eqref{sc}, which is the single copy of pure Kerr black hole spacetime. However, due to the correction of $\sigma^{\mu\nu}_{AB}$ and tetrad \eqref{tetradD0} in this order, the tensor form of the single copy still has additional terms. We obtain the composition of Maxwell field strength tensor as
\begin{equation}
        F_{\mu\nu}^{\db}= F^{\kk}_{\mu\nu}+ F^{(\dcs)}_{\mu \nu},
\end{equation}
and the corresponding correction $F^{(\dcs)}_{\mu\nu}$ has non-vanishing component
\begin{equation}
F^{(\dcs)}_{r\phi} = \frac{{\aCS}q \sin^2\theta }{12M}r^2{\dcs}'' .
\end{equation}
The corresponding sources for the zeroth copy $\scalar_{\db}$ and single copy $F_{\mu\nu}^{\db}$ are
\begin{align}
&\dn^\mu\dn_\mu {\scalar_{\db}} 
=-\frac{\ii\alpha q^2}{2M\kappa } \left(\frac{({\vtheta}/{r})}{r^2}'\right)',\\
& \dn^\mu F_{\mu\nu}^{\db}
\equiv J^{\db}_{\nu}= \left(0,0,0,   \frac{{\aCS}q\sin^2\theta}{12M}(r^2 \dcs'')'     \right).
\end{align}
Notice that the non-vanishing component can also be written as
$J^{\db}_\phi=\frac{{\aCS} q\sin^2\theta}{12M}(r^2 \dcs'')'=\partial_r F^{(\dcs)}_{r\phi }$.

\section{Petrov type I solution}
\label{section:4}
The quadratic deformation of slowly rotating Black holes in Chern-Simons modified gravity has been studied in \cite{Yagi:2012ya}, and so was the Petrov type of this solutions in their later work~\cite{Owen:2021eez}. 
We will discuss in this section how modification of Chern-Simons terms is added to the slowing rotating black hole and show non-trivial examples for the Weyl double copy in the Petrov type I spacetime.

For the Petrov type I spacetime, the Weyl scalars  $\Psi_0$ and $\Psi_4$ in \eqref{Weylspinor} vanish by careful choice of tetrad. By using principle null directions formula and the null rotation transformation in Appendix \ref{Appa}, it was found in \cite{Owen:2021eez} that the Kerr tetrad can be corrected as
\begin{equation}\begin{aligned}
\label{dCStetrad}
l^\mu\partial_\mu~=&~l^\mu_\mathrm{GR}\partial_\mu+\delta_\mathrm{CS}\partial_\phi, \\
n^\mu \partial_\mu~=&~n^\mu_\mathrm{GR}\partial_\mu  +\frac{\fm}{2}\delta_\mathrm{CS}\partial_\phi, \\
m^\mu\partial_\mu=&~m^\mu_\mathrm{GR}\partial_\mu +{\ii}\frac{ r\sin\theta}{\sqrt{2}}~\delta_\mathrm{CS}\partial_t.
\end{aligned}
\end{equation}
Here the higher order terms of $\mathcal{O}(\alpha^2a, \alpha a^2)$ have been dropped for brevity. The corresponding non-vanishing Weyl scalars are
\begin{equation}\begin{aligned}
\label{dCStetrad2}
\Psi_2^\mathrm{CS} =&~ \Psi_2^\mathrm{GR} 
=-\frac{M}{(r-{\ii}  a\cos\theta)^3},  \\
\Psi_1^\mathrm{CS}=&~ -\frac{2}{\fm}\Psi_3^\mathrm{CS} = -\frac{\ii3\sqrt{2}}{2}\frac{M}{r^2} {\sin\theta}\delta_\mathrm{CS} \,, 
\end{aligned}
\end{equation}
in which, $\delta_\mathrm{CS}$ represents
\begin{align}
  &\delta_{\mathrm{CS}}=\frac{\sqrt{1407}}{112}\frac{\alpha}{\sqrt{\kappa}M^2} \frac{a}{r^2}\times\\
&\bigg(1+\frac{2840}{603}\frac{M}{r}+\frac{64660}{4221}\frac{M^2}{r^2}+\frac{1740}{67}\frac{M^3}{r^3}+\frac{1980}{67}\frac{M^4}{r^4}\bigg)^{\frac{1}{2}}.\nonumber
\end{align}

\subsection{The decomposition}
Before we start to build double copy structure for Chern-Simons modified gravity, we just have to retrospect to specific works on general type double copy. There have been some researches that has developed a new point of view of how we treat classical double copy for the Petrov type I
spacetime~\cite{White:2020sfn, Chacon:2021wbr}.

Now we take a quick look at the Weyl spinor of Petrov type I spacetime. With the special choice of the tetrad in \eqref{dCStetrad}, the Weyl spinor reads
\begin{align}
  \Psi_{ABCD}^{[\mathrm{I}]}
=&-4\Psi_1^\mathrm{CS}o_{(A}\iota_B\iota_C\iota_{C)}+6\Psi_2^\mathrm{CS}o_{(A}o_B\iota_C\iota_{D)}\nonumber\\
&-4\Psi_3^\mathrm{CS}o_{(A}o_Bo_C\iota_{D)}.  
\end{align}
With simple algebraic insights,
we can decompose the Weyl spinor above into:
\begin{align}\label{typeI0}
\Psi_{ABCD}^{[\mathrm{I}]}&=\frac{1}{\scalar_{[0]}} {f}^\fa_{(AB} {f}^\fb_{CD)},
\end{align}
where the two Maxwell spinors are assumed to be
\begin{equation}
\begin{aligned}\label{maxwell2}
{f}^\fa_{AB}&= a_0*o_{(A}\iota_{B)}+a_1*\iota_{(A}\iota_{B)} ,\\
{f}^\fb_{CD}&=  b_0*o_{(C}\iota_{D)}
+b_1*o_{(C}o_{D)} .
\end{aligned}
\end{equation}
Whenever each parts of right-hand side of \eqref{maxwell2} are multiplied together, the product also carries symmetrization. So there is no need to worry about commutation between the spinors. What is left to deal with are some simple algebra, then polynomial equations from \eqref{typeI0} read:
\begin{equation}
\begin{aligned}
    a_0 *b_0+a_1*b_1&=6\Psi_2^\mathrm{CS}\scalar_{[0]},\\
    b_0*a_1&=-4\Psi_1^\mathrm{CS}\scalar_{[0]},\\
    a_0*b_1&=-4\Psi_3^\mathrm{CS}\scalar_{[0]}.
\end{aligned}
\end{equation}
Up to the first order of $\delta_{\mathrm{CS}}$, we can find the solutions reveal the corrections added by Chern-Simons term:
\begin{equation}
  \begin{aligned}\label{scalar0}
\scalar_{[0]}&=\scalar_{0}=-\frac{q^2}{6M}\frac{1}{r-{\ii}  a\cos\theta},\\
a_0&=b_0=
\frac{q}{(r-{\ii}  a\cos\theta)^2},\\
a_1&=-\frac{2}{\fm}b_1=-\frac{\ii q\sqrt{2}\sin{\theta}\delta_{CS}}{r}.
\end{aligned}
\end{equation}

The Maxwell spinors in \eqref{maxwell2} are solved as
\begin{equation}\label{two fields}
\begin{aligned}   
{f}^\fa_{AB}&=f^{\kk}_{AB}
-\frac{{\ii}q\sqrt{2}\sin{\theta}\delta_{CS}}{r} \iota_{(A}\iota_{B)},\\
{f}^\fb_{CD}&=f^{\kk}_{CD}
+\frac{{\ii}q\sqrt{2}\sin{\theta}\delta_{CS}}{r} \frac{\fm}{2}o_{(C}o_{D)},
\end{aligned}
\end{equation}
where $f^{\kk}_{AB}$ is the spinning Maxwell charge as shown in \eqref{sc}.
As we suspected, due to the consideration of Chern-Simons modified gravity, so the Maxwell spinors should also be altered.

The Maxwell spinor with those parameters \eqref{scalar0} can also be denoted as
\begin{equation}
\begin{aligned}\label{maxwell3}
f^{[0]}_{AB}&= a_0*o_{(A}\iota_{B)},\\
f^{[a_1]}_{AB}&=a_1*\iota_{(A}\iota_{B)},\\
f^{[b_1]}_{AB} &=b_1*o_{(A}o_{B)}.
\end{aligned}
\end{equation}
Thus, through the generalized Weyl double copy, we have found the more general type of electromagnetic fields, though they show differences at spinorial level.

Here the leading order Maxwell spinor $f^{[0]}_{AB}$ is same to the initial Kerr set up \eqref{sc} as below:
\begin{align}
f^{[0]}_{AB}= f^{\kk}_{AB} &=\frac{q}{(r-{\ii}  a\cos\theta)^2}o_{(A}\iota_{B)}.
\end{align}
After covering the spinor into tensor form, we obtain:
\begin{equation}
\begin{aligned}
    F^{[0]}_{\mu\nu}&=F^{\kk}_{\mu\nu}+F^{(\delta)}_{\mu\nu}.
\end{aligned}
\end{equation}
With the correction of the tetrad in \eqref{dCStetrad}, the non-vanishing component  of $F^{(\delta)}_{\mu\nu}$ is 
\begin{equation}
F^{(\delta)}_{r\phi}=-q\sin^2{\theta}\delta_{CS}.
\end{equation}
From \eqref{maxwell3},
the Maxwell field strength tensors $F^{[a_1]}_{\mu\nu}$ and $F^{[b_1]}_{\mu\nu}$ can also be obtained, and the non-vanishing components are:
\begin{equation}
\begin{aligned}
F^{[a_1]}_{r\phi}&=F^{[b_1]}_{r\phi}=q \sin^2{\theta} \delta_{CS},\\
F^{[a_1]}_{t\phi}&=-F^{[b_1]}_{t\phi}=q\fm \sin^2{\theta} \delta_{CS}.
    \end{aligned}
\end{equation}
While the rest of the components are identical to zero.
Notice that there is a sign difference between their components $F^{[a_1]}_{t\phi}$ and $F^{[b_1]}_{t\phi}$.

With these electromagnetic field strength tensors, which are obtained from spinor basis through the spinor-tensor transformation, we can derive the source currents as below
\begin{equation}
\begin{aligned}
&J^{[0]}_{\nu}\equiv\dn^\mu {F}^{[0]}_{\mu\nu}=(0,0,0,-q\sin^2{\theta}\delta^{'}_{CS}),\\
&J^{[a_1]}_{\nu}\equiv\dn^\mu {F}^{[a_1]}_{\mu\nu}=(0,0,0,q\sin^2{\theta}\delta^{'}_{CS}),\\
&J^{[b_1]}_{\nu}\equiv\dn^\mu {F}^{[b_1]}_{\mu\nu}=(0,0,0,q\sin^2{\theta}\delta^{'}_{CS}).
\end{aligned}
\end{equation}
When we treat those three Maxwell field strength tensors into the form \eqref{two fields}, it is clear the total sources vanish due to the opposite signatures
\begin{align}
&\dn^\mu {F}^\fa_{\mu\nu}=
\dn^\mu {F}^\fb_{\mu\nu}=0.
\end{align}
The scalar field in \eqref{scalar0} is also source-less, subject to \begin{equation}
\dn^{\mu}\dn_{\mu}\scalar_{[0]}=0.
\end{equation}
Here the covariant differential operator $D^{\mu}$ is the covariant derivative in rotational flat spacetime.

In the leading order, the sources of Maxwell fields has been splitted into two parts while they are dealt with separately as three fields. What is unforeseen is that the field strength tensor is a bit different from what is set in slowly rotating charge, that is because of the certain order we set. The corrected information such as $\Psi_1^\mathrm{CS}$ and $\Psi_3^\mathrm{CS}$ in \eqref{dCStetrad2} will only be harvest through high order ansatz, so that the higher order terms are cutoff and the low order information is preserved. From initial settings, we can deduce that metric is the principal key to determine whether the derived electromagnetic field is sourced or not.
The corrections to slowly rotating charge spinors in \eqref{two fields} are counteracted by the new added ones, and it can be interpreted as the ``neutralization'' contained in the process.

\section{Conclusion}
\label{section:5}

In this paper, we studied the Weyl double copy for slowly rotating black hole with small Chern-Simons correction. Based on the basic concept of the double copy formula and the slowly rotating black hole solutions in  Chern-Simons modified gravity, we take a step further to extend the Weyl double copy from the classical Einstein gravity to Chern-Simons modified gravity, and verify the plausibility of special case examples. Based on the sourced double copy structure, we obtain the slowly rotating charge solutions with an external source from the slowly rotating black hole solution in Chern-Simons modified gravity.

For the Petrov type D perturbative black hole solution in the Chern-Simons modified gravity, we find an additional correction of the electromagnetic field strength tensor obtained through the double copy relation. The correction appears at the intersection of the radial and azimuth coordinates, which is consistent with the correction of the metric.

For the Petrov type I perturbative black hole solution in the Chern-Simons modified gravity, we find that the additional electromagnetic filed strength tensors have the same exogenous properties at specific orders through the double copy relation, while the Maxwell equation of the total electromagnetic field remains source-less at the leading order. This indicates that metric might be the main factor that ensure whether the Maxwell equations carry source or not.

\section*{Acknowledgement}
This work is supported by the National Key Research and Development Program of China (No. 2023YFC2206200),
the National Natural Science Foundation of China (No.12375059) and the Fundamental Research Funds for the Central Universities.

\appendix
\section{Weyl scalar formalism}\label{Appa}
The derivation of Weyl scalar has alrady been examined in section \ref{spinor formula},
which can be used to have a better understanding of the spacetime. There is another technique we have used in this paper, which is the null rotation transformation, which helps us to transform Weyl scalars such that $\Psi_{0}=\Psi_{4}=0$ in the case of type I solutions. In general, the null transformation consists of three branches~\cite{Stephani:2003tm,Owen:2021eez}.

Class I: $l$ is fixed, and the rest of the tetrads are rotated as
\begin{equation}
\begin{aligned}
&l\rightarrow l,\\
&n\rightarrow n+\bar{A}m+A\bar{m}+A\bar{A}l,\\
&m\rightarrow m+Al,\\
&\bar{m}\rightarrow\bar{m}+\bar{Al}.
\end{aligned}
\end{equation}
Here $A$ represents a complex rotation coefficient, and $\bar{A}$ stands for the complex conjugation of $A$.
In this case, Weyl scalars are also transformed as
\begin{equation}
\begin{aligned}
&\Psi_{0} \rightarrow\Psi_0,  \\
&\Psi_{1} \rightarrow\Psi_{1}+\bar{A}\Psi_{0},  \\
&\Psi_{2} \rightarrow\Psi_2+2\bar{A}\Psi_1+\bar{A}^2\Psi_0,  \\
&\Psi_{3} \rightarrow\Psi_3+3\bar{A}\Psi_2+3\bar{A}^2\Psi_1+\bar{A}^3\Psi_0,  \\
&\Psi_{4} \rightarrow\Psi_4+4\bar{A}\Psi_3+6\bar{A}^2\Psi_2 +4\bar{A}^3\Psi_1+\bar{A}^4\Psi_0. 
\end{aligned}
\end{equation}

Class II: $n$ is fixed, and the rest of the tetrads are rotated as
\begin{equation}
\begin{aligned}
&\begin{aligned}l\rightarrow l+\bar{B}m+B\bar{m}+B\bar{B}n,\end{aligned} \\
&n\rightarrow n, \\
&m\rightarrow m+Bl, \\
&\bar{m}\rightarrow\bar{m}+\bar{B}l.
\end{aligned}
\end{equation}
Here $B$ represents a complex rotation coefficient, and the corresponding rotation for Weyl scalar reads
\begin{equation}
\begin{aligned}
&\Psi_0\rightarrow \Psi_0+4B\Psi_1+6B^2\Psi_2+4B^3\Psi_3+B^4\Psi_4,\\
&\Psi_{1} \rightarrow \Psi_1+3B\Psi_2+3B^2\Psi_3+B^3\Psi_4,  \\
&\Psi_{2} \rightarrow \Psi_2+2B\Psi_3+B^2\Psi_4,  \\
&\Psi_3 \rightarrow \Psi_3+B \Psi_4,  \\
&\Psi_4 \rightarrow \Psi_4. 
\end{aligned}
\end{equation}

Class III:  $l$ and $n$ can be scaled with $Y$, while the rest of the tetrads rotate in the complex plane with phase $X$:
\begin{equation}
\begin{aligned}
&l \rightarrow l/Y,\\
&n \rightarrow Yn,\\
&m \rightarrow e^{{\ii} X}m,\\
&\bar{m} \rightarrow e^{-{\ii} X}\bar{m}.
\end{aligned}
\end{equation}
The Weyl scalars are transformed as
\begin{equation}
\begin{aligned}
&\Psi_{0} \rightarrow Y^{-2}e^{2{\ii} X}\Psi_{0},\\
&\Psi_{1} \rightarrow Y^{-1}e^{{\ii} X}\Psi_1,\\
&\Psi_2 \rightarrow \Psi_2,\\
&\Psi_3 \rightarrow Ye^{-{\ii} X}\Psi_3,\\
&\Psi_{4} \rightarrow Y^2e^{-2{\ii} X}\Psi_4. 
\end{aligned}
\end{equation}


\vspace{20pt}
\begin{thebibliography}{100}

\footnotesize

\bibitem{Kawai:1985xq}
H.~Kawai, D.~C.~Lewellen and S.~H.~H.~Tye,
``A Relation Between Tree Amplitudes of Closed and Open Strings,''
Nucl. Phys. B \textbf{269}, 1-23 (1986)
doi:10.1016/0550-3213(86)90362-7

\bibitem{Bern:2008qj}
Z.~Bern, J.~J.~M.~Carrasco and H.~Johansson,
``New Relations for Gauge-Theory Amplitudes,''
Phys. Rev. D \textbf{78}, 085011 (2008)
doi:10.1103/PhysRevD.78.085011
[arXiv:0805.3993 [hep-ph]].

\bibitem{Bern:2010ue}
Z.~Bern, J.~J.~M.~Carrasco and H.~Johansson,
``Perturbative Quantum Gravity as a Double Copy of Gauge Theory,''
Phys. Rev. Lett. \textbf{105}, 061602 (2010)
doi:10.1103/PhysRevLett.105.061602
[arXiv:1004.0476 [hep-th]].



\bibitem{Bern:1999ji}
Z.~Bern and A.~K.~Grant,
``Perturbative gravity from QCD amplitudes,''
Phys. Lett. B \textbf{457}, 23-32 (1999)
doi:10.1016/S0370-2693(99)00524-9
[arXiv:hep-th/9904026 [hep-th]].

\bibitem{Saotome:2012vy}
R.~Saotome and R.~Akhoury,
``Relationship Between Gravity and Gauge Scattering in the High Energy Limit,''
JHEP \textbf{01}, 123 (2013)
doi:10.1007/JHEP01(2013)123
[arXiv:1210.8111 [hep-th]].

\bibitem{Neill:2013wsa}
D.~Neill and I.~Z.~Rothstein,
``Classical Space-Times from the S Matrix,''
Nucl. Phys. B \textbf{877}, 177-189 (2013)
doi:10.1016/j.nuclphysb.2013.09.007
[arXiv:1304.7263 [hep-th]].

\bibitem{Monteiro:2014cda}
R.~Monteiro, D.~O'Connell and C.~D.~White,
``Black holes and the double copy,''
JHEP \textbf{12}, 056 (2014)
doi:10.1007/JHEP12(2014)056
[arXiv:1410.0239 [hep-th]].

\bibitem{Adamo:2022dcm}
T.~Adamo, J.~J.~M.~Carrasco, M.~Carrillo-Gonz\'alez, M.~Chiodaroli, H.~Elvang, H.~Johansson, D.~O'Connell, R.~Roiban and O.~Schlotterer,
``Snowmass White Paper: the Double Copy and its Applications,''
[arXiv:2204.06547 [hep-th]].

\bibitem{Luna:2015paa}
A.~Luna, R.~Monteiro, D.~O'Connell and C.~D.~White,
``The classical double copy for Taub\textendash{}NUT spacetime,''
Phys. Lett. B \textbf{750}, 272-277 (2015)
doi:10.1016/j.physletb.2015.09.021
[arXiv:1507.01869 [hep-th]].

\bibitem{Berman:2018hwd}
D.~S.~Berman, E.~Chac\'on, A.~Luna and C.~D.~White,
``The self-dual classical double copy, and the Eguchi-Hanson instanton,''
JHEP \textbf{01}, 107 (2019)
doi:10.1007/JHEP01(2019)107
[arXiv:1809.04063 [hep-th]].


\bibitem{Alkac:2022tvc}
G.~Alkac, M.~K.~Gumus and M.~A.~Olpak,
``Generalized black holes in 3D Kerr-Schild double copy,''
Phys. Rev. D \textbf{106}, no.2, 026013 (2022)
doi:10.1103/PhysRevD.106.026013
[arXiv:2205.08503 [hep-th]].

\bibitem{Dempsey:2022sls}
R.~Dempsey and P.~Weck,
``Compactifying the Kerr-Schild double copy,''
JHEP \textbf{05}, 198 (2023)
doi:10.1007/JHEP05(2023)198
[arXiv:2211.14327 [hep-th]].

\bibitem{Easson:2023dbk}
D.~A.~Easson, G.~Herczeg, T.~Manton and M.~Pezzelle,
``Isometries and the double copy,''
JHEP \textbf{09}, 162 (2023)
doi:10.1007/JHEP09(2023)162
[arXiv:2306.13687 [gr-qc]].

\bibitem{Farnsworth:2023mff}
K.~Farnsworth, M.~L.~Graesser and G.~Herczeg,
``Double Kerr-Schild spacetimes and the Newman-Penrose map,''
JHEP \textbf{10}, 010 (2023)
doi:10.1007/JHEP10(2023)010
[arXiv:2306.16445 [hep-th]].

\bibitem{Ortaggio:2023cdz}
M.~Ortaggio, V.~Pravda and A.~Pravdova,
``Kerr-Schild double copy for Kundt spacetimes of any dimension,''
JHEP \textbf{2024}, no.02, 069 (2024)
doi:10.1007/JHEP02(2024)069
[arXiv:2312.00706 [gr-qc]].

\bibitem{Ceresole:2023wxg}
A.~Ceresole, T.~Damour, A.~Nagar and P.~Rettegno,
``Double copy, Kerr-Schild gauges and the Effective-One-Body formalism,''
[arXiv:2312.01478 [gr-qc]].

\bibitem{Stephani:2003tm}
H.~Stephani, D.~Kramer, M.~A.~H.~MacCallum, C.~Hoenselaers and E.~Herlt,
``Exact solutions of Einstein's field equations,''
Cambridge Univ. Press, 2003,
ISBN 978-0-521-46702-5, 978-0-511-05917-9
doi:10.1017/CBO9780511535185



\bibitem{Luna:2018dpt}
A.~Luna, R.~Monteiro, I.~Nicholson and D.~O'Connell,
``Type D Spacetimes and the Weyl Double Copy,''
Class. Quant. Grav. \textbf{36}, 065003 (2019)
doi:10.1088/1361-6382/ab03e6
[arXiv:1810.08183 [hep-th]].

\bibitem{Luna:2016due}
A.~Luna, R.~Monteiro, I.~Nicholson, D.~O'Connell and C.~D.~White,
``The double copy: Bremsstrahlung and accelerating black holes,''
JHEP \textbf{06}, 023 (2016)
doi:10.1007/JHEP06(2016)023
[arXiv:1603.05737 [hep-th]].

\bibitem{Goldberger:2016iau}
W.~D.~Goldberger and A.~K.~Ridgway,
``Radiation and the classical double copy for color charges,''
Phys. Rev. D \textbf{95}, no.12, 125010 (2017)
doi:10.1103/PhysRevD.95.125010
[arXiv:1611.03493 [hep-th]].

\bibitem{Luna:2016hge}
A.~Luna, R.~Monteiro, I.~Nicholson, A.~Ochirov, D.~O'Connell, N.~Westerberg and C.~D.~White,
``Perturbative spacetimes from Yang-Mills theory,''
JHEP \textbf{04}, 069 (2017)
doi:10.1007/JHEP04(2017)069
[arXiv:1611.07508 [hep-th]].

\bibitem{Arkani-Hamed:2019ymq}
N.~Arkani-Hamed, Y.~t.~Huang and D.~O'Connell,
``Kerr black holes as elementary particles,''
JHEP \textbf{01}, 046 (2020)
doi:10.1007/JHEP01(2020)046
[arXiv:1906.10100 [hep-th]].

\bibitem{Huang:2019cja}
Y.~T.~Huang, U.~Kol and D.~O'Connell,
``Double copy of electric-magnetic duality,''
Phys. Rev. D \textbf{102}, no.4, 046005 (2020)
doi:10.1103/PhysRevD.102.046005
[arXiv:1911.06318 [hep-th]].

\bibitem{Kim:2019jwm}
K.~Kim, K.~Lee, R.~Monteiro, I.~Nicholson and D.~Peinador Veiga,
``The Classical Double Copy of a Point Charge,''
JHEP \textbf{02}, 046 (2020)
doi:10.1007/JHEP02(2020)046
[arXiv:1912.02177 [hep-th]].

\bibitem{Luna:2020adi}
A.~Luna, S.~Nagy and C.~White,
``The convolutional double copy: a case study with a point,''
JHEP \textbf{09}, 062 (2020)
doi:10.1007/JHEP09(2020)062
[arXiv:2004.11254 [hep-th]].

\bibitem{Cristofoli:2020hnk}
A.~Cristofoli,
``Gravitational shock waves and scattering amplitudes,''
JHEP \textbf{11}, 160 (2020)
doi:10.1007/JHEP11(2020)160
[arXiv:2006.08283 [hep-th]].

\bibitem{Monteiro:2020plf}
R.~Monteiro, D.~O'Connell, D.~Peinador Veiga and M.~Sergola,
``Classical solutions and their double copy in split signature,''
JHEP \textbf{05}, 268 (2021)
doi:10.1007/JHEP05(2021)268
[arXiv:2012.11190 [hep-th]].

\bibitem{Guevara:2020xjx}
A.~Guevara, B.~Maybee, A.~Ochirov, D.~O'connell and J.~Vines,
``A worldsheet for Kerr,''
JHEP \textbf{03}, 201 (2021)
doi:10.1007/JHEP03(2021)201
[arXiv:2012.11570 [hep-th]].

\bibitem{Cardoso:2016ngt}
G.~L.~Cardoso, S.~Nagy and S.~Nampuri,
``A double copy for $ \mathcal{N}=2 $ supergravity: a linearised tale told on-shell,''
JHEP \textbf{10}, 127 (2016)
doi:10.1007/JHEP10(2016)127
[arXiv:1609.05022 [hep-th]].

\bibitem{Cardoso:2016amd}
G.~Cardoso, S.~Nagy and S.~Nampuri,
``Multi-centered $ \mathcal{N}=2 $ BPS black holes: a double copy description,''
JHEP \textbf{04}, 037 (2017)
doi:10.1007/JHEP04(2017)037
[arXiv:1611.04409 [hep-th]].

\bibitem{Cheung:2016say}
C.~Cheung and G.~N.~Remmen,
``Twofold Symmetries of the Pure Gravity Action,''
JHEP \textbf{01}, 104 (2017)
doi:10.1007/JHEP01(2017)104
[arXiv:1612.03927 [hep-th]].

\bibitem{Cheung:2017kzx}
C.~Cheung and G.~N.~Remmen,
``Hidden Simplicity of the Gravity Action,''
JHEP \textbf{09}, 002 (2017)
doi:10.1007/JHEP09(2017)002
[arXiv:1705.00626 [hep-th]].

\bibitem{Goldberger:2017frp}
W.~D.~Goldberger, S.~G.~Prabhu and J.~O.~Thompson,
``Classical gluon and graviton radiation from the bi-adjoint scalar double copy,''
Phys. Rev. D \textbf{96}, no.6, 065009 (2017)
doi:10.1103/PhysRevD.96.065009
[arXiv:1705.09263 [hep-th]].

\bibitem{Luna:2017dtq}
A.~Luna, I.~Nicholson, D.~O'Connell and C.~D.~White,
``Inelastic Black Hole Scattering from Charged Scalar Amplitudes,''
JHEP \textbf{03}, 044 (2018)
doi:10.1007/JHEP03(2018)044
[arXiv:1711.03901 [hep-th]].

\bibitem{Goldberger:2017vcg}
W.~D.~Goldberger and A.~K.~Ridgway,
``Bound states and the classical double copy,''
Phys. Rev. D \textbf{97}, no.8, 085019 (2018)
doi:10.1103/PhysRevD.97.085019
[arXiv:1711.09493 [hep-th]].

\bibitem{Chester:2017vcz}
D.~Chester,
``Radiative double copy for Einstein-Yang-Mills theory,''
Phys. Rev. D \textbf{97}, no.8, 084025 (2018)
doi:10.1103/PhysRevD.97.084025
[arXiv:1712.08684 [hep-th]].

\bibitem{Goldberger:2017ogt}
W.~D.~Goldberger, J.~Li and S.~G.~Prabhu,
``Spinning particles, axion radiation, and the classical double copy,''
Phys. Rev. D \textbf{97}, no.10, 105018 (2018)
doi:10.1103/PhysRevD.97.105018
[arXiv:1712.09250 [hep-th]].

\bibitem{Li:2018qap}
J.~Li and S.~G.~Prabhu,
``Gravitational radiation from the classical spinning double copy,''
Phys. Rev. D \textbf{97}, no.10, 105019 (2018)
doi:10.1103/PhysRevD.97.105019
[arXiv:1803.02405 [hep-th]].

\bibitem{Shen:2018ebu}
C.~H.~Shen,
``Gravitational Radiation from Color-Kinematics Duality,''
JHEP \textbf{11}, 162 (2018)
doi:10.1007/JHEP11(2018)162
[arXiv:1806.07388 [hep-th]].

\bibitem{Plefka:2018dpa}
J.~Plefka, J.~Steinhoff and W.~Wormsbecher,
``Effective action of dilaton gravity as the classical double copy of Yang-Mills theory,''
Phys. Rev. D \textbf{99}, no.2, 024021 (2019)
doi:10.1103/PhysRevD.99.024021
[arXiv:1807.09859 [hep-th]].

\bibitem{Mizera:2018jbh}
S.~Mizera and B.~Skrzypek,
``Perturbiner Methods for Effective Field Theories and the Double Copy,''
JHEP \textbf{10}, 018 (2018)
doi:10.1007/JHEP10(2018)018
[arXiv:1809.02096 [hep-th]].

\bibitem{CarrilloGonzalez:2018ejf}
M.~Carrillo Gonz\'alez, R.~Penco and M.~Trodden,
``Radiation of scalar modes and the classical double copy,''
JHEP \textbf{11}, 065 (2018)
doi:10.1007/JHEP11(2018)065
[arXiv:1809.04611 [hep-th]].


\bibitem{Keeler:2020rcv}
C.~Keeler, T.~Manton and N.~Monga,
``From Navier-Stokes to Maxwell via Einstein,''
JHEP \textbf{08}, 147 (2020)
doi:10.1007/JHEP08(2020)147
[arXiv:2005.04242 [hep-th]].

\bibitem{Cai:2013uye}
R.~G.~Cai, L.~Li, Q.~Yang and Y.~L.~Zhang,
``Petrov type I Condition and Dual Fluid Dynamics,''
JHEP \textbf{04}, 118 (2013)
doi:10.1007/JHEP04(2013)118
[arXiv:1302.2016 [hep-th]].



\bibitem{Godazgar:2020zbv}
H.~Godazgar, M.~Godazgar, R.~Monteiro, D.~Peinador Veiga and C.~N.~Pope,
``Weyl Double Copy for Gravitational Waves,''
Phys. Rev. Lett. \textbf{126}, no.10, 101103 (2021)
doi:10.1103/PhysRevLett.126.101103
[arXiv:2010.02925 [hep-th]].


\bibitem{White:2020sfn}
C.~D.~White,
``Twistorial Foundation for the Classical Double Copy,''
Phys. Rev. Lett. \textbf{126}, no.6, 061602 (2021)
doi:10.1103/PhysRevLett.126.061602
[arXiv:2012.02479 [hep-th]].

\bibitem{Chacon:2021wbr}
E.~Chac\'on, S.~Nagy and C.~D.~White,
``The Weyl double copy from twistor space,''
JHEP \textbf{05}, 2239 (2021)
doi:10.1007/JHEP05(2021)239
[arXiv:2103.16441 [hep-th]].

\bibitem{CarrilloGonzalez:2022ggn}
M.~Carrillo Gonz\'alez, W.~T.~Emond, N.~Moynihan, J.~Rumbutis and C.~D.~White,
``Mini-twistors and the Cotton double copy,''
JHEP \textbf{03} (2023), 177
doi:10.1007/JHEP03(2023)177
[arXiv:2212.04783 [hep-th]].

\bibitem{Han:2022ubu}
S.~Han,
``Weyl double copy and massless free-fields in curved spacetimes,''
Class. Quant. Grav. \textbf{39}, no.22, 225009 (2022)
doi:10.1088/1361-6382/ac96c2
[arXiv:2204.01907 [gr-qc]].

\bibitem{Han:2022mze}
S.~Han,
``The Weyl double copy in vacuum spacetimes with a cosmological constant,''
JHEP \textbf{09}, 238 (2022)
doi:10.1007/JHEP09(2022)238
[arXiv:2205.08654 [gr-qc]].

\bibitem{Bahjat-Abbas:2017htu}
N.~Bahjat-Abbas, A.~Luna and C.~D.~White,
``The Kerr-Schild double copy in curved spacetime,''
JHEP \textbf{12}, 004 (2017)
doi:10.1007/JHEP12(2017)004
[arXiv:1710.01953 [hep-th]].

\bibitem{Ball:2023xnr}
A.~Ball, A.~Bencke, Y.~Chen and A.~Volovich,
``Hidden symmetry in the double copy,''
JHEP \textbf{10}, 085 (2023)
doi:10.1007/JHEP10(2023)085
[arXiv:2307.01338 [hep-th]].


\bibitem{Godazgar:2021iae}
H.~Godazgar, M.~Godazgar, R.~Monteiro, D.~Peinador Veiga and C.~N.~Pope,
``Asymptotic Weyl double copy,''
JHEP \textbf{11}, 126 (2021)
doi:10.1007/JHEP11(2021)126
[arXiv:2109.07866 [hep-th]].

\bibitem{Mao:2023yle}
P.~Mao and W.~Zhao,
``Asymptotic Weyl double copy in Newman-Penrose formalism,''
JHEP \textbf{02}, 171 (2024)
doi:10.1007/JHEP02(2024)171
[arXiv:2312.17160 [hep-th]].

\bibitem{Easson:2020esh}
D.~A.~Easson, C.~Keeler and T.~Manton,
``Classical double copy of nonsingular black holes,''
Phys. Rev. D \textbf{102}, no.8, 086015 (2020)
doi:10.1103/PhysRevD.102.086015
[arXiv:2007.16186 [gr-qc]].

\bibitem{Chawla:2023bsu}
S.~Chawla and C.~Keeler,
``Black hole horizons from the double copy,''
Class. Quant. Grav. \textbf{40}, no.22, 225004 (2023)
doi:10.1088/1361-6382/acfe57
[arXiv:2306.02417 [hep-th]].

\bibitem{Lin:2022jrp}
G.~Lin and G.~Yang,
``Double copy for tree-level form factors. Part I. Foundations,''
JHEP \textbf{02}, 012 (2024)
doi:10.1007/JHEP02(2024)012
[arXiv:2211.01386 [hep-th]].

\bibitem{Lin:2023rwe}
G.~Lin and G.~Yang,
``Double copy for tree-level form factors. Part II. Generalizations and special topics,''
JHEP \textbf{02}, 013 (2024)
doi:10.1007/JHEP02(2024)013
[arXiv:2306.04672 [hep-th]].

\bibitem{Alkac:2023glx}
G.~Alkac, M.~K.~Gumus, O.~Kasikci, M.~A.~Olpak and M.~Tek,
``Regularized Weyl double copy,''
[arXiv:2310.06048 [hep-th]].

\bibitem{He:2023iew}
J.~L.~He and J.~H.~Huang,
``Cosmological horizons from classical double copy,''
Phys. Lett. B \textbf{851}, 138579 (2024)
doi:10.1016/j.physletb.2024.138579
[arXiv:2312.00972 [hep-th]].

\bibitem{Adamo:2014baa}
T.~Adamo and E.~T.~Newman,
``The Kerr-Newman metric: A Review,''
Scholarpedia \textbf{9}, 31791 (2014)
doi:10.4249/scholarpedia.31791
[arXiv:1410.6626 [gr-qc]].

\bibitem{Alexander:2009tp}
S.~Alexander and N.~Yunes,
``Chern-Simons Modified General Relativity,''
Phys. Rept. \textbf{480}, 1-55 (2009)
doi:10.1016/j.physrep.2009.07.002
[arXiv:0907.2562 [hep-th]].

\bibitem{Yunes:2009hc}
N.~Yunes and F.~Pretorius,
``Dynamical Chern-Simons Modified Gravity. I. Spinning Black Holes in the Slow-Rotation Approximation,''
Phys. Rev. D \textbf{79}, 084043 (2009)
doi:10.1103/PhysRevD.79.084043
[arXiv:0902.4669 [gr-qc]].

\bibitem{Easson:2021asd}
D.~A.~Easson, T.~Manton and A.~Svesko,
``Sources in the Weyl Double Copy,''
Phys. Rev. Lett. \textbf{127}, no.27, 271101 (2021)
doi:10.1103/PhysRevLett.127.271101
[arXiv:2110.02293 [gr-qc]].

\bibitem{Easson:2022zoh}
D.~A.~Easson, T.~Manton and A.~Svesko,
``Einstein-Maxwell theory and the Weyl double copy,''
Phys. Rev. D \textbf{107}, no.4, 044063 (2023)
doi:10.1103/PhysRevD.107.044063
[arXiv:2210.16339 [gr-qc]].


\bibitem{Penrose:1985bww}
R.~Penrose and W.~Rindler,
``Spinors and Space-Time,''
Cambridge Univ. Press, 2011,
ISBN 978-0-521-33707-6, 978-0-511-86766-8, 978-0-521-33707-6
doi:10.1017/CBO9780511564048

\bibitem{Penrose:1986ca}
R.~Penrose and W.~Rindler,
``SPINORS AND SPACE-TIME. VOL. 2: SPINOR AND TWISTOR METHODS IN SPACE-TIME GEOMETRY,''
Cambridge University Press, 1988,
ISBN 978-0-521-34786-0, 978-0-511-86842-9
doi:10.1017/CBO9780511524486

\bibitem{Penrose:1960eq}
R.~Penrose,
``A Spinor approach to general relativity,''
Annals Phys. \textbf{10}, 171-201 (1960)
doi:10.1016/0003-4916(60)90021-X

\bibitem{Yagi:2012ya}
K.~Yagi, N.~Yunes and T.~Tanaka,
``Slowly Rotating Black Holes in Dynamical Chern-Simons Gravity: Deformation Quadratic in the Spin,''
Phys. Rev. D \textbf{86}, 044037 (2012)
[erratum: Phys. Rev. D \textbf{89}, 049902 (2014)]
doi:10.1103/PhysRevD.86.044037
[arXiv:1206.6130 [gr-qc]].

\bibitem{Owen:2021eez}
C.~B.~Owen, N.~Yunes and H.~Witek,
``Petrov type, principal null directions, and Killing tensors of slowly rotating black holes in quadratic gravity,''
Phys. Rev. D \textbf{103}, no.12, 124057 (2021)
doi:10.1103/PhysRevD.103.124057
[arXiv:2103.15891 [gr-qc]].


\bibitem{Newman:1965tw}
E.~T.~Newman and A.~I.~Janis,
``Note on the Kerr spinning particle metric,''
J. Math. Phys. \textbf{6} (1965), 915-917
doi:10.1063/1.1704350
\end{thebibliography}
\end{document}